\documentstyle[12pt]{article}
\begin{document}

\author{Victor Novozhilov\footnote{E-mail : vnovozhilov@mail.ru}
 and Yuri Novozhilov\footnote{E-mail  : yunovo@pobox.spbu.ru} \\
V.Fock Department of Theoretical Physics, \\
St.Petersburg State University, 198504, St.Petersburg, Russia}
\title{Colour chiral solitons in low energy QCD}

\date{}
\maketitle

\begin{abstract}
We derive an effective colour chiral action with a background gauge field.
The action describes configurations of soliton-skyrmion type. Kinetic term
constant $f_0^2$ , analogue of $f_\pi ^2$ , is a phenomenological
dimensional parameter of the model; $d=4$ terms are unique up to the choice
of background gauge field. The case of SU(2) colour group is discussed in
detail. We study an isolated configuration, i.e. a configuration in a
background field which is the vacuum field forming the gluon condensate.
Thereby we introduce the condensate energy as a second parameter and scale.
Compared with the case of flavor skyrmion configuration, the colour chiral
action contains a piece with slowly decreasing terms coming from the
background vacuum field. Asymptotic behaviour at large distances shows
exponential decrease for the case of chromomagnetic condensates and periodic
otherwise . This defines a stability region for a colour soliton. The mass
is given by the positive definite integrand for a soliton of purely
bosonization origin. Contribution from the Yang-Mills action of the
background colour field has the sign opposite to bosonization part. The
baryon number current is not influenced by the background field and leads to
the standard baryon number $B=N_F/N_C$ . For $B=1/3$ we evaluated the
estimate from above $M\approx 460$ $Mev$ .
\end{abstract}


\subsection{Introduction}

In low energy QCD the ideas of Skyrmions \cite{ Skyrme,Bala,Witten,Manton}
and gauged solitons \cite{LD,Niemi} showed themself fruitful. To find stable
colour configurations can be an important next step towards understanding of
diquarks and exotic hadrons. A possibility of colour chiral solitons with
baryon number $N_F/N_C$ was mentioned in the first paper on colour
bosonization \cite{AAA+YuVN} . However, the effective action in bosonization 
\cite{AAA+YuVN} was implicitly gauge dependent, the choice of background
colour fields was not discussed, and soliton stability was not investigated.
It was found \cite{YaF} that direct application of an effective bosonized
lagrangian \cite{AAA+YuVN} does not lead to stable configurations. The idea
of colour skyrmions from different viewpoints was explored \cite
{Kaplan,Gomel,Karliner1, Karliner2} in attempt to construct a constituent
quark for $N_F=1$, but further development in this direction was suspended
after the conclusion \cite{ Karliner3} that stable colour solitons do not
exist in four dimensions (while it is possible in two dimensions, Ref.11 ).

The aim of this paper is to study a gauge invariant effective chiral
lagrangian and investigate stability of soliton configurations in the
gluonic vacuum.

The chiral colour bosonization in QCD follows , in general, the lines of
flavor bosonization \cite{AAA,Karchev,LMP,Ball} . However, the vector colour
(gluonic) field is a dynamical gauge field, while the flavor field in the
Dirac lagrangian is an external one. In order to get a chiral colour action,
the background field should be also chirally rotated giving an additional
contribution to the standard chiral action. In flavor bosonization no such
terms are present.

For a colour soliton, the background field describes soliton environment and
produces corresponding interaction terms in the effective chiral lagrangian.
In this paper we consider a separate (free) soliton, and, therefore, take
colour vacuum field as a background field. Such colour vacuum fields form
the gluon condensate. Experimentally, the condensate is positive, so that
the vacuum field is chromomagnetic. Its value will be a parameter of the
theory. The vacuum field gives rise to terms in the effective chiral
lagrangian with $R^2\sin ^2F$ and $R^4\sin ^4F$ where $R$ is a distance from
the center of the standard hedgehog configuration of the shape $F.$ .
Asymptotic behaviour of soliton configuration at $R\rightarrow \infty $ is
determined by the condensate: it is exponentially decreasing for positive
condensate and periodic otherwise.

In section 2 we derive a general expression for an effective colour chiral
action starting from the \ Dirac lagrangian with $N_C$ colours and $N_F$
flavors. In section 3 we make choice of a background field as a vacuum
colour field which forms the gluon condensate, work out an effective chiral
action in the case of the gauge group SU(2) for the hedghog configuration,
study asymptotic behavior and evaluate the mass.

\subsection{Colour bosonization and effective action}

Bosonization is a standard prescription for introducing chiral field, and
integration of chiral anomaly is usually invoked, as a way to the chiral
action. In the case of the colour chiral field the Dirac lagrangian is not
chiral invariant because of quark-gluon interaction term, so that there is
no anomaly. However, what is really important in order to get the chiral
action, is non-invariance of the fermion vacuum functional under chiral
transformations, while the classical lagrangian does not need to be
invariant. Such a non-invariance is related to the fact that the chiral
field belongs to fermionic degrees of freedom (chiral phase) present in the
functional measure, and an effective action arises from the Jacobian of the
transformation to new variables including colour chiral phase. We present
briefly the way to an effective action from this point of view \cite{Chaos}.

To illustrate the approach we consider a field $\Phi (x)$ with the
Lagrangian $L(x)$ and vacuum functional 
$$
Z=\int {\it D}\Phi e^{i\int Ldx} \eqno(1) 
$$
Let $\Pi (x)$ be a local field with quantum numbers of a composite system.
made out of $\Phi $ .The functional measure ${\it D}\Phi $ is over all
independent degrees of freedom of $\Phi $; thus it includes also degrees of
freedom of $\Pi (x)$ together with remaining degrees $X$ which are
considered inessential. If we change variables 
$$
\{\Phi \}\rightarrow \{\Pi ,X\}, \eqno(2) 
$$
and transform $Z$ 
$$
Z=\int {\it D}\Pi {\it D}XJ\left( \partial \Phi /\partial \Pi ,\partial \Phi
/\partial X\right) e^{i\int L\left( \Pi ,X\right) dx}\equiv Z_{inv}\int {\it %
D}\Pi e^{i\int L_{eff}\left( \Pi \right) dx} \eqno(3) 
$$
then after integrating out inessential variables $X$ we arrive at the
effective Lagrangian $L_{eff}(\Pi )$ for a collective variable $\Pi $
describing a composite field. The functional $Z_{inv}$ does not depend on $%
\Pi $ ; $J$ is a Jacobian of transformation $\{\Phi \}\rightarrow \{\Pi ,X\}$%
.

In practice, to find directly variables $X$ and the Jacobian is a difficult
task. However, there are some classes of collective variables $\Pi $, when
knowledge of $X$ is not necessary in order to find an effective Lagrangian $%
L_{eff}\left( \Pi \right) $ , namely, when $\Pi $ are parameters of
non-invariance groups and different classes correspond to different groups.
''Non-invariance'' is understood in relation to the vacuum functional : $%
\delta Z/\delta \Pi \neq 0$ .

Consider a group ${\it H}$ of transformations $U=\exp i\Pi $ of field $\Phi $%
, $\Phi \rightarrow \Phi ^U,U_3=U_2U_1$ , with the invariant measure ${\it D}%
(U^{\prime }U)={\it D}U$ and the vacuum functional 
$$
Z\left[ U\right] =\int {\it D}\Phi \exp \left( i\int dxL\left( \Phi
^U\right) \right) \eqno(4) 
$$
Integrating over $U$ we get ${\it H}$ - invariants 
$$
Z_0=\int {\it D}UZ\left[ U\right] ,Z_{inv}^{-1}=\int {\it D}UZ^{-1}\left[
U\right] \eqno(5) 
$$
which can be used in order to substract from $Z$ a ${\it H}$-invariant part
leaving a functional $Z_U$ for $U$ 
$$
ZZ_0^{-1}\simeq ZZ_{inv}^{-1}=Z_U\equiv \int {\it D}U\exp \left( i\int
dxL_{eff}\left( U\right) \right) \eqno(6) 
$$
and identifying an effective action for $U$ as 
$$
W_{eff}\left( U\right) =\int dxL_{eff}\left( U\right) =-i\ln \frac Z{Z\left[
U\right] } \eqno(7) 
$$
We use $\simeq $ to show that $Z_0$ and $Z_{inv}$ differ by ${\it {H}}$
-invariant terms. Thus, $L_{eff}$ is defined up to $U$ -independent terms.

We see that inessential variables $X$ were effectively integrated out in
integration over all initial degrees of freedom. A replacement $\Phi
\rightarrow \Phi ^U$ in both the measure and the lagrangian in $Z$ is just a
change of notations and cannot change $Z.$ Thus , the Jacobian $J$ in (3) is
expressed in terms of $W_{eff}\left( U\right) $.

Let us consider the Dirac lagrangian $L_\psi (G,A)$with background gluon
field $G_\mu $ and external colour field $A_\mu $ in the group SU(N)$\times
SU(N)$ 
$$
L_\psi =i\overline{\psi }\left( \widehat{\partial }+\widehat{G} +\gamma _5%
\widehat{A}\right) \psi =\overline{\psi }D\left( G,A\right) \psi ,\eqno(8) 
$$
A chiral field $U$ is defined by the following transformation of Dirac
fermions 
$$
\psi ^U=\frac 12\left[ \left( 1-\gamma _5\right) U+\left( 1+\gamma _5\right)
\right] \psi ,U^{+}U=1\eqno(9) 
$$
The quark Lagrangian $L_\psi \left( G,A\right) $ remains invariant, if
fields $G_\mu ,A_\mu $ are transformed appropriately 
$$
L_\psi \left( G,A\right) =\overline{\psi ^U}D\left( G^U,A^U\right) \psi ^U%
\eqno(10) 
$$
where $U$ -transformed fields are given by 
$$
G_\mu ^U+A_\mu ^U=U\left( G_\mu +A_\mu \right) U^{-1}+U\partial _\mu
U^{-1},G_\mu ^U-A_\mu ^U=G_\mu -A_\mu \eqno(11) 
$$
Repeated transformation with the chiral field $U_1$ gives $(G_\mu
^U)^{U_1}=G_\mu ^{U_1U}$. These fields are not symmetrical /antisymmetrical
with respect to left-right exchange. However, they are gauge transforms of
vector and axial vector fields $\widetilde{G_\mu },\widetilde{A}_\mu $ with
the gauge function $\chi $ which is square root of $U$ 
$$
G_\mu ^U=\chi \widetilde{G}_\mu \chi ^{-1}+\chi \partial _\mu \chi
^{-1},A_\mu ^U=\chi \widetilde{A}_\mu \chi ^{-1},\chi ^2=U\eqno(12) 
$$
Under a gauge transformation $g$ the function $\chi $ transforms as $\chi
^{\prime }=g\chi g^{-1}$ .

The infinitesimal chiral transformation $g_5=1+\gamma _5\lambda $ acts in
the following way 
$$
\delta G_\mu = \left[ A_\mu ,\lambda \right] ,\delta A_\mu =D_\mu \lambda 
$$
$$
\delta \psi = \gamma _5\lambda \psi ,\delta \overline{\psi }= \overline{\psi 
}\gamma _5\lambda ,\delta U=-\left( U\lambda + \lambda U\right) \eqno(13) 
$$
An important property of $U$ -transformed fields $G^U,A^U,\psi ^U, \overline{%
\psi ^U}$is that for them the chiral transformation $g_5$ is a non-chiral
gauge transformation 
$$
\delta G_\mu ^U = -D_\mu \left( G^U\right) \lambda ,\delta A_\mu ^U=\left[
\lambda ,A_\mu ^U\right] 
$$
$$
\delta \psi ^U = \lambda \psi ^U,\delta \overline{\psi }^U= -\overline{\psi }%
^U\lambda \eqno(14) 
$$
It follows that the Yang-Mills Lagrangian $L_{YM}\left( G^U\right) =\frac
1{2g^2}$tr$G_{\mu \nu }^U(G^U)^{\mu \nu }$ is invariant under chiral
transformations .

In order to find an effective action for the colour chiral field $U$ we
study functional integrals 
$$
Z_\psi \left( G,A,RS\right) = \int {\it D}\overline{\psi }{\it D} \psi \exp
i\int dxL_\psi \left( G,A\right) =\exp iW\left( G,A,RS\right) 
$$
$$
Z_\psi \left( G^U,A^U,RS\right) = \int {\it D}\overline{\psi } {\it D}\psi
\exp i\int dxL_\psi \left( G^U,A^U\right) = \exp iW\left(G^U,A^U,RS\right) %
\eqno(15) 
$$
which are also specified by a Regularization Scheme $RS$ . They play the
role of quantities $Z$ and $Z^U$ in definition of an effective action $%
W_{eff}(U)$ . Thus, we obtain 
$$
W_{eff}(G,U,RS)=-i\ln \frac{Z_\psi \left( G,A,RS\right) }{Z_\psi \left(
G^U,A^U,RS\right) }\eqno(16) 
$$
The usual way to calculate effective chiral action is to find an
infinitesimal change $\delta _UW_{eff}$ (i.e.the anomaly) and integrate it
up to $U$ . We put $U=\exp \Theta $ and introduce the anomaly ${\it A}\left(
x,\Theta \right) $ 
$$
{\it A}\left( x,\Theta \right) =\frac 1i\frac{\delta \ln Z_\psi \left( \exp
\Theta \right) }{\delta \Theta }\eqno(17) 
$$
Then 
$$
W_{eff}^\psi \left( \Theta \right) =-\int d^4x\int_0^1ds{\it A}\left(
x;s\Theta \right) \Theta \left( x\right) =\int d^4xL_{eff}^\psi \left(
U\right) -W_{WZW}\eqno(18) 
$$
where the Wess-Zumino-Witten term $W_{WZW}$ describes topological properties
of $U$ and is represented by five-dimensional integral with $x_5=s$ 
$$
W_{WZW} =\frac i{96\pi ^2}\int d^5x\varepsilon _{\mu \nu \sigma \lambda \rho
}tr[j_\mu ^{-}L_{\nu \sigma }L_{\lambda \rho }+j_\mu ^{+}R_{\nu \sigma
}R_{\lambda \rho }+ 
$$
$$
\frac 12\left( j_\mu ^{-}L_{\nu \sigma }U_sR_{\lambda \rho }U_s^{-1}+ j_\mu
^{+}R_{\nu \sigma }U_s^{-1}L_{\lambda \rho }U_s\right) -i\left( L_{\mu \nu
}j_\sigma ^{-}j_\lambda ^{-}j_\rho ^{-}+R_{\mu \nu }j_\sigma ^{+}j_\lambda
^{+}j_\rho ^{+}\right) 
$$
$$
-\frac 25j_\mu ^{-}j_\nu ^{-}j_\sigma ^{-}j_\lambda ^{-}j_\rho ^{-}]\eqno(19)
$$
where the following notations are used 
$$
j_\mu ^{-}=\overline{D}_\mu U_sU_s^{-1},j_\mu ^{+}=U_s^{-1}\overline{D}_\mu
U_s,U_s=\exp s\Theta 
$$
$$
\overline{D}_\mu U_s=\partial _\mu U_s+L_\mu U_s-U_sR_\mu 
$$
and the convention $\mu ,\nu ,...=1,2,3,4,5;L_5=R_5=0$ implied. We remind
that $W_{WZW}=0$ for SU(2) gauge group.

Eliminating external colour axial fields, $A_\mu =0,$ we get the effective
chiral Lagrangian $L_{eff}^\psi \left( U\right) $ arising from integration
over fermions with $N_F$ flavors 
$$
L_{eff}\left( U\right) =N_Ftr_C\{\frac{f_0^2}4D_\mu UD^\mu U^{-1} 
$$
$$
+\frac1{192\pi ^2} \left[ \frac 12\left[ UD_\nu U^{-1},UD_\mu U^{-1}\right]
^2-(UD_\nu U^{-1}UD^\nu U^{-1})^2\right] 
$$
$$
+\frac 1{96\pi ^2}\left( [UD^\mu U^{-1},UD^\nu U^{-1}](G_{\nu \mu }+UG_{\nu
\mu }U^{-1})+G_{\mu \nu }UG^{\mu \nu }U^{-1}\right) \}-T,\eqno(20) 
$$
where the kinetic term contains a constant $f_0^2$ which is an analogue of
the pion decay constant $f_\pi ^2$ . The last term $T$ contains higher
derivatives and reflects presence of higher radial excitations, 
$$
T=\frac{N_F}{96\pi ^2}tr_CD_{\mu}^2U^{-1}D_{\nu}^2U\eqno(21) 
$$
This term should be taken into account with another degrees of freedom
relevant to excited states and can be disregarded in the ground state
problem \cite{AVManash}.

Terms $d=4$ do not depend on regularization scheme $RS$ ; the constant $%
f_0^2 $ may look different in different $RS$ , but in applications it should
be taken from phenomenology. It should be considered as a dimensional
parameter.

The effective chiral Lagrangian $L_{QCD}\left( U\right) $ arising from the
Yang-Mills lagrangians for $G_\mu $ and $G_\mu ^U$ can be written as 
$$
L_{QCD}\left( U\right) =L_{YM}\left( G\right) -L_{YM}\left( G^U\right) = 
$$
$$
=-\frac 1{2g^2}tr_C\left[ \frac 12\left( UG_{\mu \nu }U^{-1}G^{\mu \nu
}\right) +\frac 1{16}\left[ UD_\mu U^{-1},UD_\nu U^{-1}\right] ^2\right] %
\eqno(22) 
$$
It has the sign opposite to the sign at similar structures arising in
bosonization.

The baryon number current $B_\mu $ is obtained by variation of the effective
action $W_{eff}^\psi \left( U\right) $ due to the variation $G_\mu
\rightarrow G_\mu +\omega _\mu $ with a color singlet $\omega _\mu $ . Only
the topological term $W_{WZW}$ is involved thereby 
$$
B_\mu =-i\frac{\delta W_{eff}^\psi }{\delta \omega _\mu } 
$$
$$
=\frac{N_F}{24\pi ^2N_C}\varepsilon _{\mu \nu \lambda \sigma }tr\left[
UD^\nu U^{-1}UD^\lambda U^{-1}UD^\sigma U^{-1}-3G^{\nu \lambda }\left(
UD^\sigma U^{-1}-U^{-1}D^\sigma U\right) \right] \eqno(23) 
$$
$B_\mu $ is normalized to give the baryon number $B=\frac 13$ for a quark.

\subsection{Choice of background field. Action for soliton in vacuum}

Colour configurations are always associated with background colour field $%
G_\mu $ because of necessity to maintain colour gauge invariance. In this
respect, the case of colour solitons is quite different from the case of
flavor solitons, where there is no flavour gauge invariance, and the
external flavour gauge field can be eliminated from the chiral action. We
consider the colour gauge group SU(2) with antihermitian generators $T_a=%
\frac{\tau _a}{2i}$ , where $\tau _a$ are the Pauli matrices.

The background colour field should be chosen according to the problem under
consideration. Our first step is to study a single colour soliton, i.e. a
soliton in the vacuum of gluonic field. The gluonic vacuum $\Psi _0$ is
characterized by the condensate 
$$
C_g=\left( \Psi _0,\frac{g^2}{4\pi ^2}O_{\mu \nu }^aO^{\mu \nu a}\Psi
_0\right) \cong \frac{g^2}{4\pi ^2}G_{\mu \nu }^aG^{\mu \nu a}\neq 0\eqno(24)
$$
that is by the non-zero vacuum expectation value of the Yang-Mills
lagrangian for the full quantum field $O_\mu $ represented by the background
vacuum field $G_\mu $ in our approximation. According to phenomenological
descpription $C_g\succ 0$ , so that $G_\mu $ is a chromomagnetic field in
the real case of SU(3) gauge group. The vacuum field strength $G_{kl}$ in
the temporal gauge $G_0=0$ is constant up to a time independent gauge
transformation. The effective Lagrangian $L_{eff}$ is invariant under gauge
transformations of background fields and the chiral field.

We shall consider the simplest case of a chromomagnetic vacuum background
field , when it is an Abelian-type field which is a product a coordinate
vector field $V_k$ and a SU(2) color vector $n^a$ 
$$
G_k^a=V_kn^a,V_k=-\frac 12V_{kl}x_l=-\frac 12\varepsilon _{klm}x_l\nu
_mB,G_k=gG_k^a\frac{\tau _a}{2i}\eqno(25) 
$$
where $n^a$ is a constant unit vector in the colour space, $\nu _m$ is a
constant unit vector in coordinate space, $\nu _mB=\frac 12\varepsilon
_{mlk}V_{lk}$ is the vacuum chromomagnetism and $B$ is related to the
condensate $C_g=\frac{g^2}{2\pi ^2}B^2$ . In the vacuum all directions $n^a$
and $\nu _l $ are equivalent, so that it is necessary to average over them
at the end. Such a choice of vacuum field does not lead to stability
troubles in QCD; although imaginary terms were detected at one-loop level 
\cite{Savvidy}, they disappear in all-loop treatment \cite{Elizalde}.

Let us write the chiral field in the usual way 
$$
U = \exp i\left( \frac{x_a\tau _a}R\right) F\left( R\right) =\cos F+ i{\bf r}%
\sin F, \\r_ar_a = r^2=1, r_a\tau _a={\bf r},r_a=\frac{x_a}R\eqno(26) 
$$
Under a gauge transformation $S(\overrightarrow{x})$ the chiral field $U$
transforms together with the vacuum unit colour vector ${\bf n=}n_a\tau _a $
as 
$$
U^{\prime }=SUS^{+},{\bf n}^{\prime }=S{\bf n}S^{+}+S\partial _kS^{+} 
$$
and it is convenient to restrict $S(\overrightarrow{x})$ by a condition $%
\left[ S,U\right] =0.$

Main structures in the Effective Chiral Lagrangian $L_{eff}\left( U\right) $
take on the following form 
$$
D_kU=\partial _kU+g\frac{V_k}{2i}\left[ {\bf n},i{\bf r}\right] \sin F 
$$
$$
S_{kl}=\left[ UD_kU^{+},UD_lU^{+}\right] = 
$$
$$
=4g\left( \left[ \overrightarrow{{\bf n}},\overrightarrow{{\bf r}}\right]
_lV_k-\left[ \overrightarrow{{\bf n,}}\overrightarrow{{\bf r}}\right]
_kV_l\right) \frac{\sin ^2F}R+\partial _lU\partial _kU^{+}-\partial
_kU\partial _lU^{+}\eqno(27) 
$$
where $R^2=x_kx_k$ . The kinetic structure $K$ and related non -Skyrme 
term $N$ are given by 
$$
K = tr\left( D_lU^{+}D_lU\right) =2[\left( \partial _RF\right) ^2+
\frac{2\sin ^2F}{R^2}]+2gB\left[ \overleftarrow{\nu },\overleftarrow{r}\right]
\left[ \overleftarrow{n},\overleftarrow{r}\right] \sin ^2F+ $$
$$
+\frac 14g^2B^2\left[ \overleftarrow{\nu },\overleftarrow{r}\right]
^2\left[ \overrightarrow{n},\overrightarrow{r}\right] ^2R^2\sin ^2F\eqno(28)
$$

$$
N=tr\left( D_lU^{+}D_lU\right) ^2 
$$
$$
=2\left( \left( \partial _RF\right) ^2+\frac{2\sin ^2F}{R^2}\right) ^2+\frac{%
16}{15}g^4V^4\sin ^4F+\frac 83g^2(\left( \partial _RF\right) ^2+3\frac{\sin
^2F}{R^2})V^2\sin ^2F\eqno(29) 
$$
where $A^2=V_kV_k$ and we have averaged $N$ over directions ${\bf n}$ in the
SU(2) colour space putting $\overline{n_k}=0,\overline{n_k^2}=\frac 13,%
\overline{n_k^4}=\frac 15,\overline{n_k^2n_l^2}=\frac 1{15}$ .

Similarly, we average over directions $\overrightarrow{\nu }$ of field $V_k$
in space of coordinates $x_k$ and get 
$$
\overrightarrow{V}=\frac 12\left[ \overrightarrow{\nu },\overrightarrow{r}%
\right] RB,\overline{V^2}=\frac 16B^2R^2,\overline{V^4}=\frac 14\left(
1-\left( \overrightarrow{r,}\overrightarrow{\nu }\right) ^2\right)
_{av}^2B^4R^4=\frac 2{15}B^4R^4\eqno(30)
$$
It follows that the gauge field dependent part of the Skyrmion structure is
given by 
$$
trS_{kl}S_{kl}-tr(S_{kl}S_{kl})_{B=0}=\frac{32}9g^2B^2\sin ^4F\eqno(31)
$$
while a mixed part of chirally transformed Lagrangian of the background
vacuum field acquires $\sin ^2F$ 
$$
trG_{lk}UG_{lk}U^{+}=-\frac 43g^2B^2\sin ^2F,G_{lk}=\frac g{2i}B\varepsilon
_{lkt}\nu _tn,trG_{lk}G_{lk}=-g^2B^2\eqno(32)
$$
We are now able to right down the Effective Colour Static Lagrangian 
$$
L_{eff}(U,G_k)=-N_F\frac{f_0^2}4[2(\left( \partial _RF\right) ^2+2\frac{\sin
^2F}{R^2})+\frac 29g^2B^2R^2\sin ^2F]
$$

$$
-\frac{N_F}{96\pi ^2}[\left( (\partial _RF)^2+2\frac{\sin ^2F}{R^2}\right)
^2+\frac{16}{225}g^4B^4R^4\sin ^4F+
$$
$$
+\frac 29g^2B^2R^2\left( (\partial _RF)^2+3\frac{\sin ^2F}{R^2}\right) \sin
^2F]-(\frac{N_F}{48\pi ^2}-\frac 1{2g^2})\frac 23g^2B^2\sin ^2F
$$
$$
-\left( \frac{N_F}{12\pi ^2}-\frac 1{g^2}\right) \left[ 
\frac{2\sin ^2F}{R^2}\left( \partial _RF\right) ^2+\frac{\sin ^4F}{R^4}%
+\frac 29g^2B^2\sin ^4F\right] \eqno(33)
$$
where terms with $gB$ arise from vacuum background field $G_\mu $ ,while $%
1/g^2$ terms come from the Yang-Mills lagrangian of $G_\mu $ .

The Euler-Lagrange equation for (33) for the soliton function $F(R)$ has the
form 
$$
N_Ff_0^2\left[ \left( 1+\frac{g^2B^2}6R^4\right) \sin 2F-2R\partial
_RF-R^2\partial ^2F\right] + 
$$
$$
+\left( \frac{N_F}{12\pi ^2}-\frac 1{g^2}\right) \left[ \frac{\sin ^2F\sin 2F%
}{R^2}-\sin 2F(\partial _RF)^2-2\sin ^2F\partial ^2F+\frac 23g^2B^2R^2\sin
^2F\sin 2F\right] + 
$$
$$
+\frac{N_F}{24\pi ^2}\left[ \frac{2\sin ^2F\sin 2F}{R^2}-2R(\partial
_RF)^3-3R^2(\partial _RF)^2\partial _R^2F-2\partial _R^2F\sin ^2F-(\partial
_RF)^2\sin 2F\right] + 
$$
$$
+\frac{N_F}{675}g^4B^4R^6\sin ^2F\sin 2F 
$$
$$
+\frac{N_F}{216\pi ^2}g^2B^2R^4\left[ \frac{3\sin ^2F\sin 2F}{R^2}
-4\partial _RF\frac{\sin ^2F}R-\partial _R^2F\sin ^2F\right] 
$$
$$
-\frac{N_F}{432\pi ^2}g^2B^2R^4(\partial _RF)^2\sin 2F+ \frac 13\left( \frac{%
N_F}{24\pi ^2}- \frac 1{g^2}\right) g^2B^2R^2\sin 2F=0\eqno(34) 
$$

This expression for the effective action contains terms $R^4\sin ^4F$ and $%
R^2\sin ^2F$ defining the asymptotic behaviour of $\sin F$ necessary to
obtain finite static energy or mass 
$$
M=-4\pi \int dRR^2L_{eff}\left( U,G_k\right) \eqno(35) 
$$
The contribution to the mass functional $M$ from the bosonized action sums
from the kinetic term, $d=4$ terms and the contribution of the background
vacuum field. It is easely to see that this part is positive definite and
bounded from below and provides with the soliton configuration. The
contribution from the Yang-Mills lagrangian (proportianal to $1/g^2$) is
negative and can destabilize the soliton.

We introduce dimensionless variable $\rho=Ef_0R$ , then the asymptotic
behavior at large $R$ of the decreasing function $F\left( \rho \right) $ is
represented by the following equation 
$$
\partial _\rho \left[ \rho ^2\partial _\rho F\right] - 2(1+C\rho ^4)F=0, %
\eqno(36) 
$$
where the dimensionless parameter $C=\pi^2Cg/9(Ef_0)^4$ is related to the
gluon condensate $Cg=g^2B^2/2\pi^2$. The solution of the Eq.(36) is modified
Bessel functions of the second kind $K\left( \frac34,\sqrt{\frac{C}{2}}%
\rho^2\right)\sqrt\rho $ and asymptotic behavior 
$$
F\rightarrow \rho ^{-\frac32}\exp \left( - \sqrt{\frac{C}{2}}\rho ^ 2\right)
,\rho\rightarrow \infty \eqno(37) 
$$
which guarantees that the mass $M$ is finite.

At small $\rho$, as it can be expected, the soliton function $F(\rho )$
behaves near the origin $\rho =0$ in the same manner as in the Skyrme model $%
F\left( \rho \right) \approx \pi -b\rho $.

Thus, the function $F$ of the colour soliton is quite different from that of
the flavor skyrmion.

It is easy to verify that the Baryon current $B_\mu$ is not influenced by
the background vacuum field.

According to Witten analysis \cite{Witten}, statistics of the soliton is
determined by the topological term $W_{WZ}$ which is linear in time
derivatives and reflects soliton behaviour under the rotation through a $%
2\pi $ angle. In the case of $W_{WZ}$ built on chiral colour fields,
statistics is defined by the sign $\left( -1\right) ^{N_F}$ , i.e. by odd or
even number of flavors $N_F$ . Baryon number $B$ of soliton is equal to $%
B=N_F/N_C$ . Thus, the simplest colour solitons will have $B=\frac 13,
\frac23$ that corresponds to quark and diquark.

We consider a family of trial functions 
$$
F(\rho)=\pi \sqrt{\frac{1-b\rho+a\rho^2}{1+A\rho^5}} \exp(-\frac{A}{2}%
\rho^2) \eqno(38) 
$$
where coefficients a and b are variational parameters and parameter $A=\sqrt{%
2\pi^2Cg/9(Ef_0)^4}$. We also minimize the mass functional with respect to
the scale transformation $\rho \rightarrow E\rho$. The functions (38)
reflects the behaviour at the origin and large distances (37). We look for
soliton configuration with $N_F=1$. We use the value for the gluon
condensate $C_g = (350 MeV)^4$. For the case $\alpha_s=g^2/4\pi^2
\rightarrow \infty$ we find stable soliton solutions for the wide range of
the unknown phenomenological parameter $f_0 = (10 ... 60) MeV$. In the case
of finite $\alpha_s$ we use the value for $f_0 = 16 MeV$, which corresponds
to the mass scale $\Lambda_C = 100 MeV$ of the colour bosonization \cite
{AAA+YuVN}. For the $\alpha_s$ we take the value when the corresponding
terms in (33) change their signs to opposite: $M(\alpha_s \rightarrow
\infty) = 460 MeV$ , $M(\alpha_s = 12 \pi)= 362 MeV$, $M(\alpha_s =
4\pi)=120 MeV$. For $\alpha_s =3\pi$ the contribution from the last two
terms in (33) is negative and its value exceeds all positive contributions
to mass functional.

\subsection{Conclusions and discussion}

We have derived chiral colour action in background field. The action
contains two parts: gauge invariant bosonization part and QCD part arising
from the lagrangian of background field. The background field should satisfy
standard conditions of QCD in background gauge. QCD part enters with the
sign opposite to bosonization part. We applied this action to the case of
soliton configuration defined as a configuration in a vacuum background
field related to the gluon condensate, and considered the static case.
Vacuum background field ensures exponential decrease in asymptotic region
for chromomagnetic condensate, which is essential for stability of soliton
and finiteness of mass. The (renorm-invariant) condensate is considered as a
phenomenological quantity. Bosonization part of action does not depend
explicitly on the coupling constant, while the QCD part contains it in
denominator, because of renorm-invariance properties of the QCD background
gauge. Negative contribution to mass from the QCD part increases as inverse
coupling constant. Variational estimates with the trial function with proper
asymptotics behaviour and the gluon condensate $(350Mev)^4$ shows that for
the one flavor case the mass cannot be more then 460 MeV. Solitons
definitely exist in bosonization part of action. Solitons exist in complete
action, if one seriously considers perturbation results for the coupling
constant. Solitons disappear when coupling constant becomes small, and the
QCD part cancels stabilizing terms in the mass. This statement for the
complete action is of indicative character, because of the simplest
approximation used for the QCD part.

In this paper the QCD sector was treated in classical approximation, where
only background field lagrangian is retained. This could explain, why
kinetic term for the chiral field does not arise from QCD part and QCD
contribution requires very large value of $\alpha _s$ . An extension of QCD
sector to one-loop terms may clarify these points and establish relation
with Faddeev-Niemi approach.

The problem of confinement is crucial for all coloured solitons. In two
dimensions confinement of colour solitons was demonstrated in Ref.11.

Let us consider briefly main point leading to confining potential between
colour solitons in the case of $N_C=2,N_F=1$ discussed in section 3. For
such solitons, the vacuum background field plays the role of a bag. We take
two solitons $U\left( x_1\right) $ and $U\left( x_2\right) $ described by
Eq. (26) and ask what is an intersoliton potential ${\it V}\left(
x_{12}\right) $ at large distances $x_{12}$ and large $R_1,R_2$ , that
results from lagrangian (20) for two-soliton state 
$U(x_1,x_2;{\bf n,\nu })=U\left( x_1,{\bf n,\nu }\right) U\left( x_2,{\bf n,\nu }\right) $ 
after averaging over common colour and coordinate unit vectors ${\bf n,\nu }$ of
vacuum background field (25)

$$
{\it V}\left( x_{12}\right) F_1F_2=\left\langle H\left[ U\left( x_1,x_2;%
{\bf n,\nu }\right) \right] \right\rangle -\left\langle H\left[ U\left( x_1,%
{\bf n,\nu }\right) \right] \right\rangle -\left\langle H\left[ U\left( x_2,%
{\bf n,\nu }\right) \right] \right\rangle 
$$
where $\left\langle H\right\rangle $ denotes averaging over ${\bf n,\nu }$
and $H$ is the static energy density following from (20). Second and third
terms in ${\it V}$ are given by $-L_{eff}$ in (33). The QCD part of the
effective lagrangian (22) does not contribute. A gauge invariant potential
is 
\begin{eqnarray*}
{\it V}\left( x_{12}\right) F_1F_2 &=&\frac{f_0^2}4\left\langle tr\left[
DU_2DU_1^{+}+DU_1DU_2^{+}\right] \right\rangle \cong tr\{\left[ G_k\left(
x_1\right) ,U_1\right] \left[ G_k\left( x_2\right) ,U_2^{+}\right] \} \\
{\it V}\left( x_{12}\right) &=&\frac{f_0^2}{36}g^2B^2\left( \overleftarrow{%
x_1}-\overleftarrow{x_2}\right) ^2 \\
&&
\end{eqnarray*}

Such potential describes confined solitons. However, this does not explain
whether an isolated colour soliton can exist and propagate. Discussion of
confinement of colour solitons will be given in a separate paper.

\end{document}